\documentclass[cameraready]{Interspeech}
\title{Robust Audio-Visual Target Speaker Extraction \\ with Multiple Enrollment Fusion}

\author[affiliation={1,5}]{Zhan}{Jin}
\author[affiliation={1,5}]{Bang}{Zeng}
\author[affiliation={3,4}]{Peijun}{Yang}
\author[affiliation={3,4}]{Jiarong}{Du}
\author[affiliation={6}]{Wei}{Ju}
\author[affiliation={6}]{Yao}{Tian}
\author[affiliation={3,1,4}]{Juan}{Liu}
\author[affiliation={2,3,5}, correspondingauthor]{Ming}{Li}
\address{
    $^1$ School of Computer Science, Wuhan University, Wuhan, China \\
    $^2$ School of Artificial Intelligence, The Chinese University of Hong Kong, Shenzhen, China \\
    $^3$ School of Artificial Intelligence, Wuhan University, Wuhan, China \\
    $^4$ School of Cyber Science and Engineering, Wuhan University, Wuhan, China \\
    $^5$ Digital Innovation Research Center, Duke Kunshan University, Kunshan, China \\
    $^6$ AI Center, OPPO, Beijing, China
}

\email{ming.li.cuhksz@gmail.com}


\usepackage{comment}

\begin{document}

\maketitle

\begin{abstract}
    Audio-Visual Target Speaker Extraction (AVTSE) is crucial for cocktail party scenarios. Leveraging multiple cues --such as utterance-level speaker embeddings or steady face images, and frame-level lip motion or facial expression features  --can significantly improve performance. However, real-world applications often suffer from intermittent signal loss, especially for frame-level cues. This paper systematically investigates the robustness of multi-enrollment fusion under varying degrees of modality missing. Results show that while full multimodal fusion excels under ideal conditions, its performance degrades sharply when encountering unseen modalities missing during the testing. Crucially, training with a high missing rate dramatically enhances robustness, maintaining stable performance even under severe test-time modality missing. We demonstrate that fusing the complementary one frame of face image with frame-level lip features achieves both strong performance and robustness for the AVTSE task. The model and codes are shared.
\footnote{the link to the resource will be added after the blind review process}
\end{abstract}

\section{Introduction}
The “cocktail party effect” describes the human ability to focus on a specific speaker’s voice in a noisy acoustic environment. Recently, Target Speaker Extraction (TSE) has become an increasingly significant task, as it directly aims to isolate and enhance the voice of a target speaker from a multi-talker mixture using prior information. The core of this task lies in leveraging pre-enrolled cues that encapsulate the target speaker’s characteristics, which serves as a guiding signal for the extraction model to pinpoint and separate the desired voice from the background. Any information that represents the target speaker’s vocal characteristics can potentially serve as such a cue.

Various modalities are used to characterize the target speaker, broadly categorized into utterance-level and frame-level features. Early audio-visual TSE methods placed greater emphasis on frame-level lip regions \cite{gabbay2018seeing,alfouras2018conversation,wu2019time,li2024audio,li2024iianet}, as lip movements contain rich contextual information which is highly correlated with speech. Moreover, the visual modality remains unaffected by the acoustic environment, making it robust in complex auditory conditions. Gabbay \cite{gabbay2018seeing} proposed an audio-visual separation network that adopts a pre-trained VGG-Face network to extract a face descriptor for the target speaker. Afouras \cite{alfouras2018conversation} proposed using a pre-trained lip-reading model \cite{stafylakis2017combining} to extract multi-frame lip embeddings of the target speaker as visual cues to assist TSE. In USEV \cite{pan2022usev}, the leading role of lip movements in TSE tasks is analyzed under various speech overlap conditions, and in \cite{pan2022selective} self-supervised learning is applied to demonstrate the importance of synchrony between the target speaker’s visual cues and the target speech for TSE performance.

\begin{figure*}[t]
\centering
\includegraphics[width=.8\textwidth]{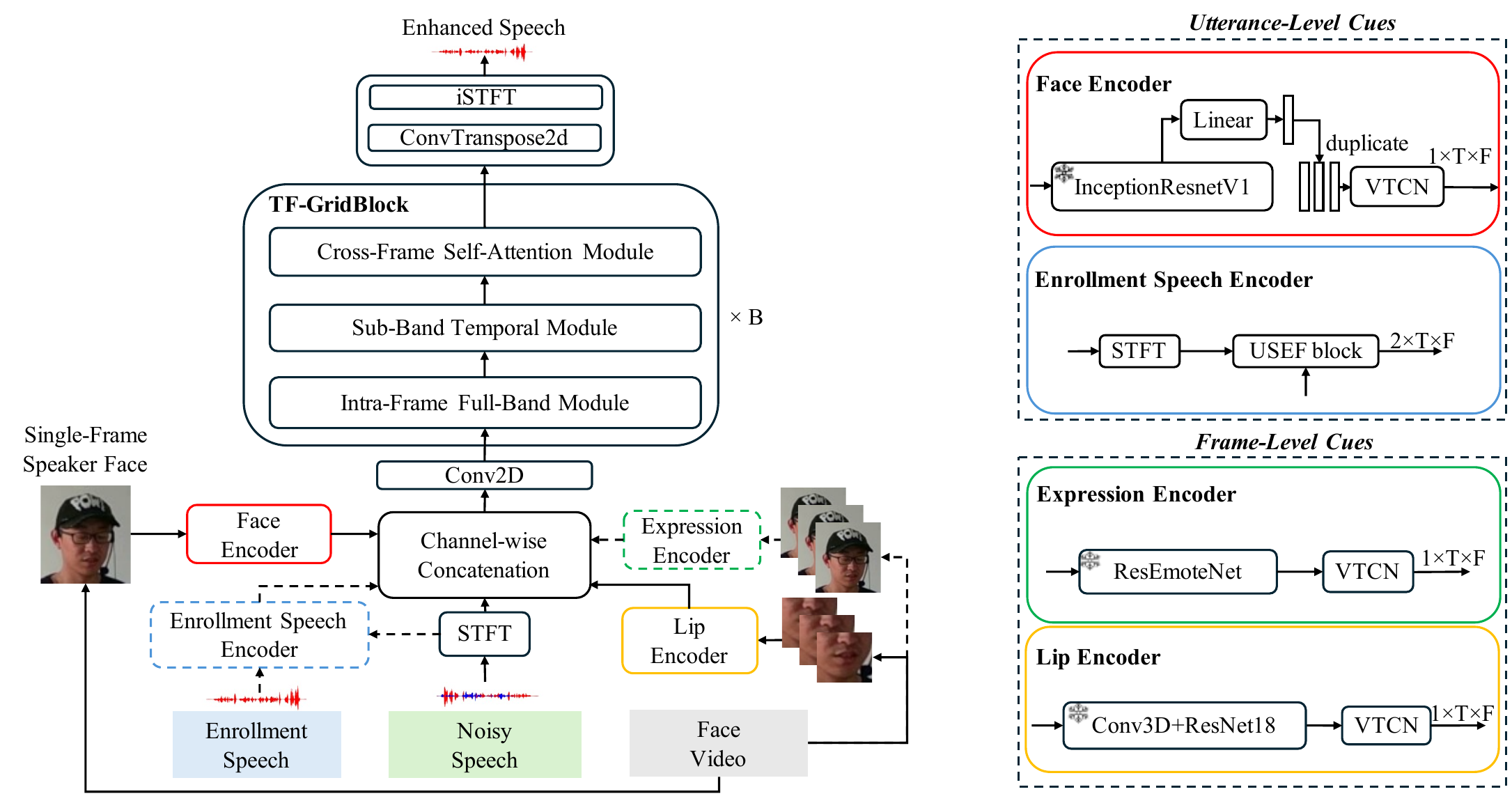}
\caption{Proposed AVTSE system structure with multiple enrollment fusion}
\label{fig:fig_1}
\end{figure*}

However, TSE methods that depend on frame-level visual information are susceptible to real-world robustness issues such as face occlusion, head movement, signal interruption, etc., which can lead to missing facial frames. This has motivated studies exploring methods to mitigate the resulting performance degradation. First, models are designed to extract useful information from available frames to compensate for missing data. ImagineNet \cite{pan2023imaginenet} employs multi-task learning combining video frame inpainting and audio-visual TSE, enhancing model performance across different video frame occlusion ratios. However, the visual encoder only utilizes lip movements, and its training and evaluation are conducted under matched conditions, either both occluded or both non-occluded.

Second, utterance-level information, including steady-face features and speaker enrollment voice, can be incorporated alongside frame-level cues. VisualVoice \cite{gao2021visualvoice} leveraged face embeddings extracted with an end-to-end face network, building a multi-task learning framework that jointly learns audio-visual TSE and cross-modal embedding distance. Two scenarios were compared where video clips are subtly misaligned and occluded during both training and test to simulate real-world conditions. Although the correlation between face and lip embeddings was analyzed, the simulated occlusion scenarios remained relatively simple, comparing only fully-occluded versus non-occluded conditions during both training and testing. Voice embeddings are commonly adopted as the reference cue in TSE \cite{wang2023tf,subakan2021attention,luo2020dual,zeng2025usef}, but they are highly sensitive to the quality of the enrollment speech and are inconvenient for speakers at first glance. Afouras proposed a two-stage model framework \cite{afouras2019my} that selectively uses visual embeddings to assist TSE and leverages the estimated speech as voice enrollment in the second stage. Through realistic simulations, model performance was empirically investigated under varying ratios of face occlusion. However, that work did not show the performance of the purposed voice+visual enrollment method when the model is trained without occlusion but tested under varying lip occlusion ratios, and it only examined the correlation between two enrollment modalities. Face images contain partial speaker information such as gender and age, providing robust utterance-level support for TSE. Moreover, they offer a convenient enrollment method for speakers at first glance. The aforementioned works did not focus on study the coupling relationships among multi-modality cues in TSE, the potential of applying multiple enrollment fusion under varying missing ratios has not been explored.

In this paper, we address the problem of partially missing face frames from two perspectives. 1. We explore a wider variety of modal information pairs within the model architecture, analyze the effects of four types of speaker cues—lip, face, expression, and enrollment speech—and investigate their correlations and functional differences in TSE tasks. 2. We employ training strategies that expose the model to severely degraded scenarios, thereby enhancing its adaptability to incomplete frames, and validate the effectiveness of these modalities under various degrees of modality missing. Section II describes the baseline architecture and embedding extraction for the four modalities, Section III discusses experimental details, and Section IV presents the conclusion.

\begin{table*}[htbp]
\centering
\caption{Performance on the AVSEC3 Test Set with 0\% occlusion during the training and 0\%, 40\% and 80\% occlusions during the test}
\label{tab:table1}
\begin{tabular}{lrrrrrrrrr}
\toprule
 & \multicolumn{3}{c}{SISDR} & \multicolumn{3}{c}{PESQ} & \multicolumn{3}{c}{STOI} \\
\cmidrule(lr){2-4} \cmidrule(lr){5-7} \cmidrule(lr){8-10}
Model & 0\% & 40\% & 80\% & 0\% & 40\% & 80\% & 0\% & 40\% & 80\% \\
\midrule
Mix & -4.860 & -4.860 & -4.860 & 1.176 & 1.176 & 1.176 & 0.615 & 0.615 & 0.615 \\
Lip & 12.221 & 7.893 & 3.043 & 2.399 & 1.912 & 1.662 & 0.874 & 0.776 & 0.672 \\
Lip-Expr & 12.561 & 8.221 & 1.693 & 2.438 & 1.890 & 1.516 & 0.879 & 0.783 & 0.651 \\
Lip-Face & 12.661 & 7.862 & 5.031 & 2.451 & 1.901 & 1.816 & 0.881 & 0.769 & 0.697 \\
Lip-Enroll Speech & 12.968 & 10.174 & 8.019 & 2.490 & 2.153 & 2.016 & 0.886 & 0.818 & 0.768 \\
Lip-Expr-Face & 12.636 & 9.206 & 6.109 & 2.460 & 2.070 & 1.942 & 0.881 & 0.800 & 0.732 \\
Lip-Expr-Face-Enroll Speech & 13.137 & 8.822 & 6.941 & 2.522 & 2.036 & 1.988 & 0.887 & 0.790 & 0.753 \\
\bottomrule
\end{tabular}
\end{table*}

\begin{table*}[htbp]
\centering
\caption{Performance on the AVSEC3 Test Set with 80\% occlusion during the training and 0\%, 40\% and 80\% occlusions during the test}
\label{tab:table2}
\begin{tabular}{lrrrrrrrrr}
\toprule
 & \multicolumn{3}{c}{SISDR} & \multicolumn{3}{c}{PESQ} & \multicolumn{3}{c}{STOI} \\
\cmidrule(lr){2-4} \cmidrule(lr){5-7} \cmidrule(lr){8-10}
Model & 0\% & 40\% & 80\% & 0\% & 40\% & 80\% & 0\% & 40\% & 80\% \\
\midrule
Mix & -4.860 & -4.860 & -4.860 & 1.176 & 1.176 & 1.176 & 0.615 & 0.615 & 0.615 \\
Lip & 12.338 & 12.151 & 12.071 & 2.425 & 2.404 & 2.312 & 0.876 & 0.871 & 0.868 \\
Lip-Expr & 12.659 & 12.427 & 12.087 & 2.464 & 2.434 & 2.404 & 0.881 & 0.875 & 0.869 \\
Lip-Face & 12.827 & 12.548 & 12.298 & 2.478 & 2.451 & 2.428 & 0.884 & 0.878 & 0.873 \\
Lip-Enroll Speech & 13.073 & 12.917 & 12.828 & 2.509 & 2.489 & 2.479 & 0.888 & 0.884 & 0.882 \\
Lip-Expr-Face & 12.846 & 12.623 & 12.242 & 2.484 & 2.463 & 2.436 & 0.885 & 0.879 & 0.872 \\
Lip-Expr-Face-Enroll Speech & 13.151 & 13.027 & 12.913 & 2.539 & 2.524 & 2.507 & 0.889 & 0.885 & 0.882 \\
\bottomrule
\end{tabular}
\end{table*}

\section{Methods}

\subsection{System overview}
The proposed system shown in Figure~\ref{fig:fig_1} consists of a speech encoder, multiple target speaker encoders, namely enrollment speech encoder, lip encoder, face encoder, expression encoder, a modal fusion block, separator blocks, and a speech decoder.

The speech encoder applies STFT to the mixed speech, producing a complex spectrogram. The modal fusion block includes attractors for different modalities and a fusion module. Except for enrollment speech embeddings, which employ a cross-attention module, all visual embeddings (lip, face, expression) first pass through an attractor with a Visual Temporal Convolutional Network (V-TCN) structure \cite{pan2023scenario}, projecting the target speaker information from different modalities to features on the same plane with a unified shape. These encoder outputs, together with the complex spectrogram, are then channel-wise concatenated and passed through a 2D convolution layer in the time-frequency direction before entering the separator blocks. This facilitates effective multiple enrollment fusion and enhances TSE performance.

The separator backbone includes six TFGridBlock modules, each comprising an intra-frame full-band module, a sub-band temporal module, and a cross-frame self-attention module \cite{wang2023tf}. The separator outputs estimated target speaker's spectrogram features in the time-frequency domain. The decoder includes a 2D transposed convolution layer followed by an iSTFT to transform the tensor back to a waveform.

\subsection{Target speaker encoders}\label{subsec:encoders}
Here we have enrollment speech encoder and face encoder at the utterance level as well as lip encoder and expression encoder at the frame-level.

\subsubsection{Enrollment speech encoder at the utterance-level}
First, both the mixed speech and the enrollment speech are transformed into time-frequency domain features via STFT. Second, enrollment embeddings are extracted using the Universal Speaker Embedding Free (USEF) block \cite{zeng2025usef}, which adds a learnable multi-head cross-attention module between the enrollment and mixed TF features. The enrollment features serve as Query, and the mixed speech features serve as Key and Value, to compute voice embeddings that fully leverage the contextual relationship between enrollment and mixed speech.

\subsubsection{Face encoder at the utterance-level}
Face embeddings are extracted using the InceptionResNetV1 model \cite{szegedy2017inception}. The enrollment input is a face image of the target speaker; in our experiment, this is simulated by randomly extracting a face image of the target speaker from the video clip. The extracted face embeddings are then duplicated along the time dimension before passing through the feature attractor.

\subsubsection{Lip encoder at the frame-level}
For the lip encoder, we use the same pre-trained ResNet-18 model as in \cite{li2024audio}, which achieves 84\% lip-reading accuracy on the LRW dataset. Lip embeddings are extracted using the frontend model, including a 3D convolution layer and four residual convolution blocks. Before passing through the attractor, linear interpolation is applied to the embeddings to unify the time dimension with that of the mixed spectrogram.

\subsubsection{Expression encoder at the frame-level}
Facial expression embeddings are extracted using ResEmoteNet \cite{roy2024resemotenet}, a facial emotion recognition model that achieves SOTA performance on multiple emotion benchmarks including FER2013 \cite{goodfellow2013challenges}, RAF-DB \cite{li2017reliable}, AffectNet-7 \cite{mollahosseini2017affectnet} and ExpW \cite{zhang2018facial}. For each frame containing the target speaker's face, we employ a frontend model pre-trained on RAF-DB to extract expression embeddings, discarding the final linear layers originally used for classification. Our hypothesis is that emotional content is present in both audio and visual modalities, and that paralinguistic information often manifests through facial expressions, thereby providing contextual cues that extend beyond identity.

\subsection{Training with missing data strategy}
We implement a modality occlusion function that allows custom occlusion ratios on speech signals of varying sequence lengths. During training, we use two different occlusion ratios—
zero and 80\%—to expose the model to both no-occlusion and heavy occlusion scenarios. For testing, we perform speech extraction under three occlusion ratios: 0\%, 40\%, and 80\%.
The occlusion function replicates the approach described in \cite{afouras2019my}, specifically zeroing out consecutive frames to represent occluded speaker cues during those periods. Each clip contains multiple randomly positioned occluded segments, with the total duration equal to the three length ratios mentioned above, and each individual segment length constrained to 15–25 frames.

\begin{table*}[t]
\centering
\caption{Performance on AVSEC3 Dev Set under different mixing scenarios according to the evaluation plan \cite{blanco2023avse}}
\label{tab:table3}
\begin{tabular}{lcccccc}
\toprule
 & \multicolumn{3}{c}{Speech+Noise} & \multicolumn{3}{c}{Speech+Speech} \\
\cmidrule(lr){2-4} \cmidrule(lr){5-7}
\textbf{System} & \textbf{SISDR} & \textbf{PESQ} & \textbf{STOI} & \textbf{SISDR} & \textbf{PESQ} & \textbf{STOI} \\
\midrule
Mix & -4.4 & 1.15 & 0.68 & -5.0 & 1.17 & 0.60 \\
AV-CrossNet\cite{kalkhorani2025av} & 14.3 & 2.75 & 0.92 & 17.3 & 3.23 & 0.95 \\
SAV-GridNet\cite{pan2023scenario} & 14.2 & 2.68 & 0.91 & 17.5 & 3.23 & 0.95 \\
Lip-Face (ours) & 14.4 & 2.57 & 0.91 & 17.5 & 3.08 & 0.95 \\
\bottomrule
\end{tabular}
\end{table*}

\begin{table}[t]
\centering
\caption{Performance on AVSEC3 Leaderboard Evaluation Set, PESQ is applied here according to the evaluation plan}
\label{tab:table4}
\begin{tabular}{lccc}
\toprule
\textbf{System} & \textbf{SISDR} & \textbf{PESQ} & \textbf{STOI} \\
\midrule
ict\_avsu\_1 & 8.989 & 3.087 & 0.881 \\
Smiip\_try2 & 12.701 & 3.004 & 0.886 \\
Lip-Face (ours) & 13.259 & 3.087 & 0.894 \\
\bottomrule
\end{tabular}
\end{table}

\section{Experimental results}
\subsection{Datasets}
The 3rd COG-MHEAR Audio-Visual Speech Enhancement Challenge (AVSEC-3) constructs challenge data from various public datasets \cite{blanco2023avse}. Two types of speech mixtures are considered: a target speaker mixed with a single interferer, or a target speaker mixed with non-speech noise. Speaker data are sourced from LRS3 \cite{afouras2018lrs3}, and non-speaker noise data from CEC1 \cite{graetzer2021clarity}, DEMAND \cite{thiemann2013demand}, DNS 2nd Version \cite{dubey2023icassp}, MedleyB Audio \cite{bittner2014medleydb}, and ESC-50 \cite{piczak2015dataset}. The AVSEC-3 Training Set comprises approximately 34,500 scenes, totaling 113 hours, with 605 target speakers; interferers are drawn from 405 competing speakers and about 7,300 noise files. The AVSEC-3 Dev Set includes around 3,300 scenes, and the AVSEC-3 Test Set contains 2,400 scenes. For the occlusion comparison experiments, we used the Test Set. To compare with other systems, we accordingly evaluate our system on the Leaderboard Evaluation Set, and on Dev Set against other SOTA systems.

\subsection{Implementation Details}
Audio data are sampled at 16 kHz. In the STFT encoder, the window size is 128, hop size is 64, yielding a complex spectrogram with 65 frequency bins. After 2D convolution fusion, the feature channels expand from the concatenated size to 128. In each TF-GridBlock, both LSTM layers are bi-directional with 240 hidden units and 128 feature units; the 1D transposed convolution layer transforms the 256 channels back to 128 before input to the next LSTM layer.
% % 
% \begin{align}
%     SI-SDR = 10 \log_{10} \left( \frac{ || \alpha \hat{s} ||^2 }{ || \alpha \hat{s} - s ||^2 } \right)
%     \label{equation:sisdr}
% \end{align}
% %

SISDR is used for training, and a hybrid loss function of SISNR and STOI is used for fintuning where the STOI loss weight is 0.5. During training, dynamic mixing is applied. The batch size is 32. The optimizer is Adam with an initial learning rate of 1e-3. The learning rate is halved if validation does not improve for 3 consecutive epochs, and training stops if validation does not improve for 10 consecutive epochs. Evaluation metrics include SISDR \cite{le2019sdr}, PESQ \cite{rix2001perceptual} and STOI \cite{taal2011algorithm}.

For each frame containing the target speaker's face, we employ a frontend model pre-trained on the RAF-DB dataset to extract facial expression embeddings, discarding the final linear layers originally used for classification. Our hypothesis is that emotional content is present in both audio and visual modalities, and that paralinguistic information often manifests through facial expressions, thereby providing contextual cues that extend beyond identity.

\subsection{Results and Analysis}
Experimental results are first compared under ideal conditions without occlusion in Table~\ref{tab:table1} to analyze the synergy among different cues, and then under three testing occlusion ratios in Table~\ref{tab:table2}, investigating the impact of modality occlusion under different training regimes. Both tables are evaluated on the challenge test set using wide-band PESQ. To compare with other systems, Table~\ref{tab:table3} shows state-of-the-art systems on the challenge dev set, and Table~\ref{tab:table4} shows the best-performing systems on the challenge leaderboard. It is noticed that Table~\ref{tab:table4} uses narrow-band PESQ according to the challenge standard metrics, unlike all other tables that apply wide-band PESQ.

\subsubsection{Complementarity of different cues under Zero Occlusion}
Under ideal conditions with no occlusion, as shown in Table~\ref{tab:table1}, the full multiple fusion system integrating all four cues achieves the best performance, confirming the complementarity that different modalities provide. Face embeddings contribute noticeable gains by offering utterance-level speaker information. However, expression embeddings do not outperform face embeddings, suggesting they may not effectively decouple dynamic emotion from static identity within the tested framework. In contrast, enrollment speech embeddings yield substantial improvements when added, highlighting the value of complementary speaker information from the acoustic domain.

\subsubsection{Impact of Occlusion under Different Training Regimes}
In Table~\ref{tab:table1}, when models are trained without occlusion, performance degrades severely as test-time occlusion increases. In this setting, utterance-level embeddings demonstrate superior robustness compared to frame-level cues, while expression embeddings are highly sensitive to occlusion. This indicates that models trained on pristine data over-rely on the availability of specific modalities.

Training with a high occlusion rate as in Table~\ref{tab:table2} fundamentally improves the robustness. Models then maintain strong and stable performance across all test occlusion levels. All modalities, including expression and face embeddings, contribute positively even under high occlusion, unlike in the zero-occlusion training scenario. This shows that high-occlusion training strategy guides the model to effectively utilize available cues without overdependence on the complete set at the frame-level.

We further demonstrate that fusing the complementary one frame of face image with frame-level lip features (our Lip+Face system)achieves both strong performance and robustness for the AVTSE task.

\subsubsection{Comparison with other systems}
To situate our findings within the broader research landscape, we compare our proposed Lip-Face system with leading systems on the AVSEC-3 benchmark, as shown in Table \ref{tab:table4}. While the system \textit{ict\_avsu\_1} achieves the highest PESQ, our Lip-Face system attains comparable PESQ and superior SISDR and STOI scores. Notably, our model's performance is competitive with the top-ranking \textit{Smiip\_try2} in SISDR and exceeds it in STOI. This demonstrates that the fusion of utterance-level face embeddings with frame-level lip cues provides a robust and effective identity representation for the TSE task, even when compared to other state-of-the-art approaches on a standard hidden test set.

Further comparison with recent state-of-the-art models AV-CrossNet and SAV-GridNet on the dev set in Table \ref{tab:table3} reveals a key trade-off. While our model's SISDR is comparable to SAV-GridNet and PESQ is slightly lower, it achieves an identical STOI score. More importantly, the primary advantage of our system lies not in maximizing ideal-condition metrics but in its engineered \textit{robustness} to modality loss. Unlike typical systems trained and evaluated solely on pristine data, our model is explicitly optimized through high-occlusion-rate training (80\%) to handle severe, real-world incomplete cues. As evidenced in Table \ref{tab:table2}, this strategy yields stable performance across all test-time occlusion levels (0\%, 40\%, 80\%), a capability not demonstrated by conventional systems. Therefore, compared to systems that excel only under ideal settings, our work prioritizes and achieves reliable performance in practical scenarios where modalities are intermittently missing. The results in Table \ref{tab:table3} and Table \ref{tab:table4} confirm that while our system remains competitive on standard benchmarks, it consistently performs well under diverse and challenging missing-modality conditions.

\subsection{Conclusion}
This study re-evaluates Multiple Enrollment
Fusion for AVTSE. Under ideal conditions, face embeddings provide complementary information of the target speaker, while facial  expression embeddings show limited added value in the current framework. Training strategy critically determines real-world robustness. Training without occlusion leads to fragility when fame-level cues are missing, whereas high-occlusion training ensures consistent performance by preventing over-reliance on any single cue. We further demonstrate that fusing the complementary one frame of face image with frame-level lip features achieves both strong performance and robustness.

% \ifcameraready
%      The Interspeech 2026 organizers
% \else
%      The authors
% \fi
% would like to thank ISCA and the organizing committees of past Interspeech conferences for their help and for kindly providing the previous version of this template.

\vfill
\pagebreak

\section{Generative AI Use Disclosure}
Generative AI tools were used for limited language editing purposes, including improving clarity and correcting grammatical issues. No substantive content, analysis, or conclusions were generated by AI. The authors remain fully responsible for the content of this manuscript.

\bibliographystyle{IEEEtran}
\bibliography{mybib}

\end{document}